\documentclass{article}
\usepackage{amsfonts}
\usepackage{lscape}

\input{tcilatex}
\begin{document}

\title{\hspace{-4cm}{\small \hspace{-4cm}} \\
\relax Split-then-Combine simplex combination and selection of forecasters.}
\author{A.S.M. Arroyo\thanks{{\small a.martin@uam.es. Universidad Aut{\'o}%
noma de Madrid (UAM): Econom{\'\i}a Cuantitativa, E-III-306 , Avda.
Francisco Tom{\'a}s y Valiente 5, 28049 Madrid (Spain).} } \and A. de Juan
Fern{\'{a}}ndez\thanks{{\small Corresponding author: aranzazu.dejuan@uam.es.
Universidad Aut{\'{o}}noma de Madrid (UAM): Econom{\'{\i}}a Cuantitativa,
E-III-307 , Avda. Francisco Tom{\'{a}}s y Valiente 5, 28049 Madrid (Spain).} 
}}
\maketitle

\begin{abstract}
{\small This paper considers the Split-Then-Combine (STC) approach (Arroyo
and de Juan, 2014) to combine forecasts inside the simplex space, the sample
space of positive weights adding up to one.} {\small As it turns out, the
simplicial statistic given by the center of the simplex compares favorably
against the fixed-weight, average forecast. Besides, we also develop a
Combine-After-Selection (CAS) method to get rid of redundant forecasters. We
apply these two approaches to make out-of-sample one-step ahead combinations
and subcombinations of forecasts for several economic variables. This
methodology is particularly useful when the sample size is smaller than the
number of forecasts, a case where other methods (e.g., Least Squares (LS) or
Principal Component Analysis (PCA)) are not applicable.}


\end{abstract}

\newpage

\section{Introduction}

Forecasters have access to a wide variety of information and forecasting
techniques, thus leading to a considerable degree of heterogeneity or
redundancy among them. A weighted average forecast is expected to perform
better than individual ones because this way we can diversify away
idyosincratic forecast misspecifications, thus reducing the variance of the
forecast. The simplest example is the (fixed weight) arithmetic average.
More sophisticated methods that make use of varying weights usually do not
improve the average in empirical applications because of the instability of
the estimated weights (a problem known as \textit{forecast combination
puzzle, Stock and Watson (2004)}); in particular, when an increasing number
of forecasts requires us to estimate an increasing number of weights (a
problem known as \textit{the curse of dimensionality}). The forecast
combination puzzle has been considered by Smith and Wallis (2009), who
pointed out that the failure of more sophisticated combination methods is
due to the estimation of the combining weights. On the other hand, Barrow
and Kourentzes (2016) found that the average forecast does not perform well
in the presence of irregular data, suggesting the use of the median forecast
as the best combination method. Similar results are found by Genre et al.
(2013).

With forecast (or model)-specific combinations, forecasting is often based
on predicting the same variable independently by forecasters. However,
analysts who are interested in forecasting a variable from a specific source
should not ignore the forecasts from other competing sources. A forecast
combination is in fact influenced by all the forecasts; hence, the
relationship among individual forecasts is lost when forecasts are
independently analyzed. Only a few methods have been suggested that
incorporate dependence between forecasts. Multivariate models could
incorporate dependence between forecasts if we knew such a dependence.
Alternatively, we can engage straightaway with weight distributions based on
given individual forecast errors, as dependence between weights can be
incorporated directly, thus increasing forecast accuracy. We follow this
idea and proposeast accuracy for economic variables.

One important difference between modeling forecast-specific combinations and
weight distributions is that weights are directly dependent on each other on
an aggregated level. Constraints on weights to have non-negative values and
sum up to one lead to spurious effects on their covariance structure. In
particular, each row or column of the variance matrix of a vector of weights
sums up to zero. Given that the variances are always positive, this implies
that some covariances are forced towards negative values.

Independent modeling and forecasting with forecast-specific combinations is
not only unattractive since it ignores dependence patterns among (relative)
weights, but also because weights often fail to be coherent in the sense of
the erratic way in which the covariance associated with two specific weights
can fluctuate in sign as we move from a full combination to lower and lower
dimensional subcombinations. In fact, there is no relationship between the
variance matrix of a subcombination and that of the full combination.
Besides, variances may display different rank orderings as we form
subcombinations, which could lead to implausible forecasts.

Also, avoided forecasts in a subcombination will result in an increase of
weights for some other forecasts. In combinations, not only are there common
elements in the denominator of two weights, but also elements common to
numerator and denominator in each weight. Avoided forecasts in a
subcombination thus affect both the numerator and the denominator, and the
dependence between forecasts is therefore not as easy to predict.

All combinations are subcombinations of a larger one. Since the covariance
between two weights depends on which other forecasts are reported in the
dataset, there is no guarantee that a plot of a subcombination exhibits
similar or even compatible patterns with the plot of the original dataset,
even if the forecasts not included in the subcombination are irrelevant
(redundant).

There is thus incoherence of the correlation between weights as a measure of
dependence. Note, however, that the ratio of two weigths remains unchanged
when we move from a full combination to a subcombination. Therefore, as long
as we work with scale invariant functions (i.e., ratios), we shall be
subcombinationally coherent.

Since standard descriptive statistics (e.g., arithmetic mean and standard
deviation) are not informative with combinations, in this paper we propose a
time-varying method to combine, select, and recombine forecasts based on
Aitchison (1982, 1986), who characterizes compositions as vectors having a
relative scale and identifies its sample space with the simplex. More
crucial than the constraining property of compositional data is the
scale-invariant property of this kind of data. Indeed, when we are
considering only few forecasts of a full combination we are not working with
constrained data but our data are still compositional. This approach has
been successfully applied to various fields; see, for instance, Billheimer
et al (2001), Egozcue and Pawlowsky-Glahn (2005), and van den Boogaart,
Tolosana, and Bren (2009), a software package available now to deal with
compositional data. To our knowledge, it has not been applied to
combinations of forecasts. Compositional Data Analysis (CODA) is a
well-established set of statistical methods for the analyses of
compositional data, defined in general as a data vector with positive
elements summing to a constant value and thereby containing only relative
information (Pawlowsky-Glahn and Buccianti, 2011). Thus, CODA enables a
coherent and correct modeling of dependent weights by recognizing the sign
and sum constraints.

Traditional decomposition techniques provide inconsistent results when
applied to compositional data as they do not recognize the implicit
constraints of summing to a constant (Aitchison, 1982, 1986):
mathematically, compositional data lie in the bounded space of the simplex
while traditional decomposition techniques are defined for data in the real
space. Aitchison (1986) showed that by making log-ratio transformations it
is possible to express compositional data in the real space where the data
can be analyzed with conventional models and then transformed back into the
simplex. We make use of the centered log-ratio (CLR) transformation to
express the weights in the real space. The CLR transformation takes the
logarithm of each weight divided by the geometric mean. This transformation
maintains the initial constraint in the weights as its elements sum to 0 by
construction but resulting values are real. The inverse CLR transformation
takes the data back to the simplex with the closure operator $\mathcal{C}$
that divides the exponential of each entry by the sum of all entries.
Aitchison (1986) also defined addition and subtraction operations obeying
conventional rules of arithmetic, maintaining the result of the operation in
the simplex. These two operations applied to vectors of weights are defined,
respectively, as the closures of the ratio and product of weights. They will
be used in our analysis and are essential when modeling weight
distributions. The difference of two vectors of weights measures the
distance between them in compositional data similarly to subtraction on the
real axis. The sum of two vectors of weights is the opposite operation and
can be compared with addition on the real axis.

The analysis that is presented in this paper uses the Split-Then-Combine
(STC) approach of Arroyo and de Juan Fern{\'{a}}ndez, 2014, to generate the
weights of a combination. Because they are restricted to be positive and sum
up to one, we propose the center of the simplex $g$\ as our basic simplicial
combination vector. To get a subcombination of forecasts, we develop a
Combination-After-Selection (CAS) procedure to recombine the best subset of
forecasts, again with positive weights adding up to one. Finally, we compare
both the full combination and the CAS subcombination with the benchmark
average forecast, that has been shown hard to beat in the literature. It is
important to note that our combination vector is just the gravity center of
the simplex whose weights are not estimated, thus avoiding the combination
puzzle.

Our analysis improves forecast-specific combinations by using a CODA-based
combination vector of weights. We find that just the simplicial mean $g$\
provides in general more accurate forecasts than the average forecast (which
is just the neutral point in the simplex) in combinations as well as
subcombinations. CODA enables coherent modeling of forecast-specific
combinations where dependences between forecasts are explicitly modeled, so
a relative improvement in the weight for one forecast leads to a decline in
the relative weight for the remaining ones. Thus, CODA models provide a more
satisfactory combination as relative dependence between forecasts is taken
into account.

The paper is organized as follows: the next section describes the STC
approach both in the Euclidean and simplex spaces. Then, we explain the CAS
strategy. In the empirical application, in section 4, we pull out
information provided by panels of quarterly periodicity from a pool of
expert forecasters for the US macroeconomy over the period 1991--2018.
Forecast accuracy of simplicial combinations are compared with the uniform
benchmark arithmetic average. Finally, some concluding remarks complete the
paper.

\section{The Split-Then-Combine (STC) approach}

Arroyo and de Juan (2014) proposed the Split-Then-Combine approach to
generate combinations across $J$ forecasts $\left( \widehat{Y}_{t,j}\right) $%
, $j=1,2,...,J$, along $t=1,2,...,T$ periods using the expression:%
\[
\widehat{Y}_{t}^{(m)}=\omega _{t,1}^{(m)}\widehat{Y}_{t,1}^{(m)}+\omega
_{t,2}^{(m)}\widehat{Y}_{t,2}^{(m)}+...+\omega _{t,J}^{(m)}\widehat{Y}%
_{t,J}^{(m)} 
\]%
where the weights $\omega _{t,j}^{(m)}$ vary in two dimensions: (1) from one
period to the next; and (2) from one panel to another. A panel is a division
of the frequency data. For example, if we are working with monthly data, we
will have 12 panels, one for each month; if we work with quarterly data, we
will have four panels, one for each quarter. Panels take into account the
different behavior of the time series among seasons, but STC can also be
applied to time series with lower frequency than quarterly or monthly data.%
\footnote{%
See Bujosa-Brun et al (2019) for an application of the STC approach to
annual data with only one panel.}

The weights of the STC approach must satisfy two restrictions: be positive
and sum up to one; the latter, to avoid biased combinations if individual
forecasts are unbiased. Arroyo and de Juan (2014) developed the STC in the
Euclidean Space. The analysis presented in this paper extends and improves
the STC in the simplex space (Aitchison, 1986).

In order to see the differences between both methods, we first briefly
review the STC approach in the Eucidean space; then, we expand the STC
approach to the simplex space.

\subsection{The STC approach in the Euclidean Space}

Table 1 shows how the STC approach works in the Euclidean space. Columns 2
to 5 show the forecasts of the variable of interest for panel $m$.\ Each
element of this column represents the forecast of each forecaster for a
given period. For instance, $\widehat{Y}_{2,1}^{(m)}$ is the forecast from
forecaster 2 for period 1 in panel $m$. The 6th column shows the cross
average by period for the $J$ forecasts; that is, $\overline{\widehat{Y}}%
_{J,1}^{(m)}$ is the average of the $J$ forecasts for the first forecasting
period. The 6th row shows the time average by forecaster, that is, $%
\overline{\widehat{Y}}_{1,T_{1}}^{(m)}$\ is the average over time of all the
forecasts from the first forecaster. Column 7 reports the real data of the
variable and the 7th row shows the precision of each forecast average with
respect to the overall average $\left( \overline{\overline{\widehat{Y}}}%
_{J,T_{1}}^{(m)}\right) $. This measure is used to construct the weights $%
\omega $ that will be assigned to each forecast in the STC approach in the
Euclidean space.\bigskip

\begin{center}
\emph{Insert Table 1 around here\bigskip }
\end{center}

The STC weights $\omega $ are then computed using the information up to time 
$T_{1}$ for each panel using the expression: 
\[
\omega _{j,T_{1}}^{(m)}=\frac{\left( \overline{\widehat{Y}}_{j,T_{1}}^{(m)}-%
\overline{\overline{\widehat{Y}}}_{J,T_{1}}^{(m)}\right) ^{-2}}{%
\dsum\limits_{j=1}^{J}\left( \overline{\widehat{Y}}_{j,T_{1}}^{(m)}-%
\overline{\overline{\widehat{Y}}}_{J,T_{1}}^{(m)}\right) ^{-2}} 
\]%
and these weights are then used to form the STC combination in $T_{1}+1$ for
panel $m$:%
\[
\widehat{Y}_{T_{1}+1}^{(m)}=\omega _{1,T_{1}}^{(m)}\widehat{Y}%
_{1,T_{1}+1}^{(m)}+\omega _{2,T_{1}}^{(m)}\widehat{Y}_{2,T_{1}+1}^{(m)}+...+%
\omega _{J,T_{1}}^{(m)}\widehat{Y}_{J,T_{1}+1}^{(m)} 
\]%
This expression must be computed for each panel, $m=1,2,...,M$. These
weights satisfy two restrictions: they are positive and add up to one. Once
we get forecasts at $T_{1}+1$, we re-compute the weights by rolling over
another one-step-ahead combination for $T_{1}+2$, and so on, always
satisfying the same two restrictions.

The unit-sum\emph{\ }and non-negativeness constraints on weights, however,
give rise to a number of known issues that make inappropriate the Euclidean
geometry. Thus, standard methods from multivariate statistics are
inapplicable for computing weigths in combinations. Among others: (i)
weights cannot be normally distributed due to the bounded range of their
values; (ii) due to the constant sum constraint, each row $i$\ of the
variance-covariance matrix of a vector of random weights adds up to zero,

\begin{center}
$\sum_{j=1}^{J}C\left[ W_{i},W_{j}\right] =C\left[ W_{i},\sum_{j=1}^{J}W_{j}%
\right] =C\left[ W_{i},1\right] =0$
\end{center}

giving rise to the singularity of these matrices; that is, the
variance-covariance matrix is orthogonal to the $J\times 1$\ vector of ones.
A classical way to get rid of singularity is to erase one weight, but
results will depend on which one is erased, not being an operation that is
permutation invariant; (iii) Since $\sum_{j\neq i}^{J}C\left[ W_{i},W_{j}%
\right] =-V\left[ W_{i}\right] <0$, some covariances between weights are
forced towards negative values, leading to negative bias. In particular,
with only two weights, $V\left[ W_{1}\right] =V\left[ W_{2}\right] $ and
their correlation is $-1$. Hence, correlations are not free to range over
the usual interval $\left( -1,+1\right) $; (iv) With negative bias, what is
the meaning of zero correlation between two weights in a combination? What
correlations between restricted weights would have arisen if the
corresponding unrestricted weights had been uncorrelated? Even when two
restricted weights are correlated, it is by no means safe to conclude that
the corresponding unrestricted weights quantities are correlated
(many-to-one function from the Euclidean space to the simplex space); (v) by
definition, not only is there a common forecast in the denominator of two
weights, but also there is a common forecast in the numerator and
denominator of each weight, resulting in spurious correlation between
weights; (vi)\emph{\ }most importantly, there is no relationship between the
variance-covariance matrix of a subcombination of weights and the
subvariance-covariance matrix of the same weigths in the full combination.
Besides, variances may display different and unrelatable rank orderings as
we form subcombinations. This is known as subcombinational incoherence. For
instance, the covariance between weight $W_{1}$ and $W_{2}$ along time in a
full combination of 4 weights may be completely different, even in sign,
from the covariance of the corresponding weights $S_{1}=W_{1}/\left(
W_{1}+W_{2}+W_{3}\right) $ and $S_{2}=W_{2}/\left( W_{1}+W_{2}+W_{3}\right) $
in a subcombination where the fourth weight is excluded. There is thus
incoherence of the correlation between weights as a measure of dependence.
Note, however, that the ratio of two components remains unchanged when we
move from full combination to a subcombination: $S_{1}/S_{2}=W_{1}/W_{2}$,
so that as long as we work with scale invariant functions (i.e.,ratios), we
shall be subcombinationally coherent; (vii) the covariance between two
weights depends on those others who have been included in the combination.
Hence, there is no guarantee that the behavior of a pair of weights in a
combination exhibits similar or even compatible patterns with that of the
same pair of weights in alternative subcombinations (and all combinations
are possible subcombinations of a larger one), even if the forecasts not
included are irrelevant (i.e., nonsense of scatterplots\ for pairs of
weights over time); (viii) finally, since the construction of
subcombinations from combinations is similar to the construction of
combinations from unrestricted weights, we may expect the same difficulties
in relating variance-covariance matrices of vectors of weights in the
simplex and those in the the Euclidean space.

All of these issues lead us to consider the simplex space as the best one to
overcome these problems.

\subsection{The STC approach in the Simplex Space}

Consider a $T\times J$ panel $\widehat{\mathbb{Y}}$ of $T$\ out-of-sample
forecasts $\widehat{Y}_{t,j}$\ produced over time by $J$\ forecasters on
some variable of interest $Y_{t}$, and $\mathbb{A}$ be its related panel of
prediction accuracies $a_{t,j}\equiv \left( \widehat{Y}_{t,j}-Y_{t}\right)
^{-2}\in \mathbb{R}_{+}$. Then, the matrix

$\mathbb{W\equiv }\left( 
\begin{array}{ccc}
w_{1,1} & ... & w_{1,J} \\ 
... & ... & ... \\ 
w_{t,1} & ... & w_{t,J} \\ 
... & ... & ... \\ 
w_{T,1} & ... & w_{T,J}%
\end{array}%
\right) \equiv \left( 
\begin{array}{c}
w_{1\bullet }^{\prime } \\ 
... \\ 
w_{t\bullet }^{\prime } \\ 
... \\ 
w_{T\bullet }^{\prime }%
\end{array}%
\right) \equiv \left( 
\begin{array}{ccccc}
w_{\bullet 1} & ... & w_{\bullet j} & ... & w_{\bullet J}%
\end{array}%
\right) $

with weigths $w_{t,j}\equiv a_{t,j}/\sum\limits_{j=1}^{J}a_{t}$ represents $T
$ combination vectors $w_{1\bullet }$, ..., $w_{T\bullet }$\ such that $%
w_{t,j}\geq 0$ for all $t$ and $j$, and $\sum\limits_{j=1}^{J}w_{t,j}=1$ for
all $t$. Thus, $w_{t\bullet }^{\prime }$ is just a $1\times J$\ point in a
simplex space $\mathbb{S}^{J-1}$ of positive weights adding up to one of
dimension $J-1$. The function $\mathcal{C}:%
\mathbb{R}
_{+}^{J}\mapsto \mathbb{S}^{J-1}$ that transforms a vector of precisions $%
a_{t\bullet }\in 
\mathbb{R}
_{+}^{J}$ into a vector of weights $w_{t\bullet }\in \mathbb{S}^{J-1}$ is
called a closure transformation $w_{t\bullet }=\mathcal{C}\left( a_{t\bullet
}\right) $. Since this operator\ cancels out any constant, $\mathcal{C}%
\left( ca_{t\bullet }\right) =\mathcal{C}\left( a_{t\bullet }\right) $, it
is scale invariant. Hence, we just need to work with scale invariant
functions (e.g., ratios or logratios). Every statement about vectors in $%
\mathbb{S}^{J-1}$ will be fully expressed in terms of logratios in $%
\displaystyle%
\mathbb{R}_{+}^{J-1}$ with inferences transformed back from $%
\displaystyle%
\mathbb{R}_{+}^{J-1}$ into combinational statements in $%
\displaystyle%
\mathbb{S}^{J-1}$. In particular, we will use the center $g$ of $\mathbb{W}$
as our benchmark simplicial statistic, which is based on the centered
logratio transformation $\mathtt{clr}:\mathbb{S}^{J-1}\mapsto \mathbb{R}$,%
\begin{equation}
x_{t,j}=\mathtt{clr}\left( w_{t,j}\right) :=\ln w_{t,j}-\frac{1}{J}%
\sum_{j=1}^{J}\ln w_{t,j}=\ln \frac{w_{t,j}}{\tprod%
\nolimits_{j=1}^{J}w_{t,j}^{1/J}}\equiv \ln \frac{w_{t,j}}{g\left(
w_{t\bullet }\right) }
\end{equation}

where $g\left( w_{t\bullet }\right) $, the geometric average of the $J$\
forecasts for the $t^{th}$\ observation, is the gravity center $g$ of $%
\mathbb{W}$. This function may be interpreted as a bijection $\mathbb{S}%
^{J-1}\leftrightarrow \mathbb{H}^{J-1}$\ between $\mathbb{S}^{J-1}$ and a
vector subspace $\mathbb{H}^{J-1}:=\left\{ x_{t\bullet }\in \mathbb{R}%
_{+}^{J}:\sum_{j=1}^{J}x_{t,j}=0\right\} $ of $\mathbb{R}_{+}^{J}$\
orthogonal to the vector of ones. The inverse clr transformation is then
defined by

\begin{equation}
\mathtt{clrInv}\left( x_{t\bullet }\right) :=\mathcal{C}\left( \exp
x_{t\bullet }\right) =\mathcal{C}\left( w_{t\bullet }/g\left( w_{t\bullet
}\right) \right) =\mathcal{C}\left( w_{t\bullet }\right) =w_{t\bullet }\in 
\mathbb{S}^{J-1}
\end{equation}

that is, $\mathtt{clrInv}$\ allows us to go from $\mathbb{R}_{+}^{J-1}$\
back to $\mathbb{S}^{J-1}$. Finally, the center $g$ of $\mathbb{W}$ is given
by

\begin{equation}
g\equiv \mathtt{clrInv}\left( \overline{x}_{T\bullet }\right) =\mathcal{C}%
\left(
\tprod\nolimits_{t=1}^{T}w_{t,1}^{1/T},...,\tprod%
\nolimits_{t=1}^{T}w_{t,J}^{1/T}\right) \equiv \mathcal{C}\left( g\left(
w_{\bullet 1}\right) ,...,g\left( w_{\bullet J}\right) \right)
\end{equation}

that is, the point in the simplex given by the closure of the geometric
averages of weights over time will be the combination vector of the STC
simplex.

While vectors of weights are subcombinationally incoherent, the ratio of two
weights remains unchanged when we move from a full combination to a
subcombination; that is, $a_{t,i}/a_{t,j}=w_{t,i}/w_{t,j}=s_{t,i}/s_{t,j}$
for all $t$. Hence, as long as we work with ratios or logratios, we shall be
subcombinationally coherent. Therefore, we only consider relative precision
among forecasts: each weight in a combination vector will have no meaning on
itself isolated from the others. In particular, the variation matrix $%
\Upsilon $ with elements given by the sample variance over time, $\Upsilon
_{i,j}\equiv var\left( \ln \frac{w_{\bullet i}}{w_{\bullet j}}\right) $,
with diagonal elements all $0$, will be used to define the total variation
in $\mathbb{W}$ as $\upsilon ^{2}:=\sum_{i=1}^{J-1}\sum_{j=i+1}^{J}\Upsilon
_{i,j}$. Then, $\upsilon $\ will be a proper measure of distance among
forecasts in cluster analysis, from perfect association ($\upsilon \text{{}}%
=0$) to perfect independence ($\upsilon =+\infty $).

\section{Combination-after-Selection (CAS)}

The combination after selection procedure looks for those forecasts that
best or orthogonally contribute to improve the simplex STC full combination $%
g\in \mathbb{S}^{J-1}$. Thus, we preselect from $g\in \mathbb{S}^{J-1}$
those forecasts whose weights $\left( w_{1},...,w_{I}\right) $\ are greater
than the benchmark average $\mathcal{C}\left( \mathbf{1}_{J}\right) =1/J$
weight, the neutral point in the simplex. Then, we convert this subvector
into the CAS subcombination $\mathcal{C}\left( w_{1},...,w_{I}\right)
=\left( s_{1},...,s_{I}\right) \in \mathbb{S}^{I-1}$\ inside a simplex of a
lower dimension $I-1$ so that $s_{1}>0,...,s_{I}>0$ and $s_{1}+...+s_{I}=1$.
Sometimes, especially when $J>>T$, we perform another subsequent selection
by choosing those forecasts from the CAS preselection that are orthogonal to
each other, avoiding this way redundant forecasts. To do this, we have also
carried out cluster and biplot (Gabriel (1971) analyses.

A selected CAS subcombination $\mathcal{C}\mathbb{S}:\mathbb{S}^{J-1}\mapsto 
\mathbb{S}^{I-1}$\ will be viewed as taking place in two stages: a selection
of $I<J$\ forecasts by a selecting $%
\displaystyle%
I\times J$\ matrix $\mathbb{S}$,\ followed by its closure,

\begin{equation}
\mathcal{C}\mathbb{S}\left( g\right) =\mathcal{C}\left( \mathbb{S}g\right) :=%
\frac{\left( w_{1},...,w_{I}\right) ^{\prime }}{w_{1}+...+w_{I}}=\left(
s_{1},...,s_{I}\right) ^{\prime }
\end{equation}

For $I=3$, the CAS subcombination can be represented in a ternary digram by
barycentric coordinates (height of the point over the side of the triangle
opposite to it). Similarly, for $I=4$, it can be represented by a
tetrahedron where each possible 3-forecast subcombination vector is found by
projecting every 4-forecast vector onto the side opposite to the vertex
corresponding to the removed forecast.

\subsection{CAS from clusters of forecasts}

Forecasts are seldom homogeneous. Often there are several subgroups,
corresponding to a different, unknown subpopulation, with a distinct
behavior. In order to find possible subgroups of forecasts, we apply a
hierarchical algorithm of agglomeration with two steps: first, we consider
all forecasts as isolated groups; then, we proceed upwards using two
clustering criteria:

\begin{enumerate}
\item The ward criteria: Harmonic\textbf{\ }weighted average distance\textbf{%
\ }between forecasts in a cluster.

\item The complete criteria: Maximum distance\textbf{\ }between forecasts in
a cluster.
\end{enumerate}

We define redundant forecasts those who belong to the same cluster. Our CAS
subcombination of clusters is formed by first combining the forecasts in
each cluster, then by finding the simplicial center of all clusters. 

The dendrogram contains information on the marginal distribution of each
coordinate (althoug, it does not contain information on the relationship
between coordinates). Each coordinate is represented in the horizontal axis.
The vertical bar going up from each one of these coordinate axes represents
the variance of that specific coordinate, and the contact point is the
coordinate mean.

\subsection{CAS from a biplot of forecasts}

To better visualize the structure of $\mathbb{W}$, we just need to center it
through a translation by the inverse of its center:

\begin{equation}
\mathbb{W}_{c}\equiv \left( \frac{\left( w_{t,j}/g\left( w_{\bullet
j}\right) \right) _{j=1,...,J}}{\sum_{h=1}^{J}w_{t,h}/g\left( w_{\bullet
h}\right) }\right) _{t=1,...,T}
\end{equation}

This has also the effect of moving the center $g$\ of $\mathbb{W}$ to the
benchmark average $\mathcal{C}\left( \mathbf{1}_{J}\right) $. Moreover, if
we scale $\mathbb{W}_{c}$ by powering to $\upsilon ^{-1}$, we obtain a
combination matrix with unit total variance, but with the same relative
contribution of each logratio in the variation matrix. 

The biplot represents simultaneously the rows and columns of a centered $%
T\times J$ matrix by means of a rank-2 approximation which, in the least
squares sense, is provided by the singular value decomposition of $%
\displaystyle%
\mathbb{W}_{c}$. Observations are represented as dots and forecasts as
arrows from the center of the plot. Redundant forecasts, lying on a common
line, will show a one-dimensional\emph{\ }pattern. 

\subsection{Selection strategy}

The CAS approach that selects forecasts from the center $g$ of $\mathbb{W}$
can be summarized in the following steps:

\begin{enumerate}
\item Given a $T\times J$\ table $\widehat{\mathbb{Y}}$\ of $J$\ forecasters
over $T$\ time periods in a given season (month or quarter in our cases)%
\footnote{%
Once again, this procedure can also be applied to data with lower
periodicity than monthly or quarterly data. See for example, Bujosa et al
(2019) for an application of the simplicial methods proposed in this paper
using annual data.}, compute the related $T\times J$\ table $\mathbb{A}$\ of 
$1\times J$\ vectors $a_{t\bullet }^{\prime }$ of prediction accuracies for
each time period $t\in \left[ 1,T\right] $.

\item Convert $\mathbb{A}$ into a $T\times J$\ table $\mathbb{W}$ of
combination vectors $w_{t\bullet }^{\prime }$ of weights inside the simplex;
that is, weights in each row of $\mathbb{W}$ are positive and add up to one.

\item For each $t\in \left[ 1,T\right] $, compute a $1\times J$\ vector of
logweights in differences with respect to its average over $J$ using the 
\texttt{clr} transformation.

\item Calculate the gravity center $g$ of $\mathbb{W}$ as a combination
vector of forecasts.

\item Select the CAS subcombination of those forecasts with simplicial
weights larger than $1/J$. Where appropriate, when $J>>T$, make another
subselection from the previous CAS by applying cluster and biplot analyses
to find non-reduntant forecasts with orthonormal coordinates.

\item Go back to the simplex with the \texttt{clrInv} transformation.

\item Repeat steps 1-6 for all panels.

\item Generate rolling, out-of-sample, one-step-ahead combination vectors to
forecast next year's corresponding seasons and compare them to their
realized value.
\end{enumerate}

\section{Empirical application}

We apply the $STC$ in the Simplex and $CAS$\ combination procedures to the
variables defined in table 2, where we include their definition and the
samples used to form the combinations of forecasts. Here, we deal with
forecasts obtained from the Survey of Professional Forecasters (SPF) from
the Federal Reserve Bank of Philadelphia. Blanks in the Survey due to the
entry and exit of forecasters are fulfilled following the same strategy as
in Poncela et al (2011), that is, we only consider one-step-ahead forecasts
and select only those forecasters whithout missing data. When there is a
missing datum, we use the two-steps-ahead forecast to fill it. Forecasters
with more than four consecutive missing data are excluded. For each sample,
we only take into account balanced panels. This strategy is also used in
Lahiri, Peng and Zhao (2015). Because of the entry and exit of forecasters
in the survey, we also analyze different sample sizes, depending on the
number of included forecasters. In table 3, we show, for each variable, the
number of forecasters chosen in each subsample. The combinations of
forecasts are computed for the periods 2015 to 2018. Note that, in some
samples, the number of forecasts is larger than the number of observations,
a fact that cannot be treated with other methods (e.g., regression and
PCA).\bigskip

\begin{center}
\emph{Insert tables 2 and 3 around here}\bigskip
\end{center}

To analyze the prediction accuracy of combinations, we look at four
well-known measures : Mean Error (ME), Root Mean Squared Error (RMSE), Mean
Absolute Percentage Error (MAPE), and Median Absolute Percentage Error
(MdAPE). Although in general these measures produce similar results, there
are some differences depending on the type of the combination considered.

We compute three kind of combinations: two with varying weights ($STC$ in
the simplex, $S\_STC$ and $CAS$) and the fixed-weight arithmetic average
combination ($AVE$)\footnote{%
We consider as benchmark the average because our objective functions are
symmetric. As suggested by a referee, we also examine the Median as our
objective function. The results, however, are very similar and are available
upon request.}.

\subsection{General results}

\begin{center}
\emph{Insert table 4 around here}

\bigskip
\end{center}

We have analyzed 1266 values of accuracy measures. General results are shown
in table 4. According to the type of weights, they favored fixed weights in
469 cases (37\%) and varying weights in 797 (63\%). With respect to the
latter, 194 (15.3\%) favored S\_STC\ and 603 (47.6\%) $CAS$. In general, a
time-varying simplicial combination outperforms the fixed weights; in
particular, the S\_STC\ gives is best in 15.3\% of cases while $CAS$\ is in
47.6\%. Based on this overall result, we may conclude that selecting
forecasts improve the combination for the period 2015 - 2018.

Although in general simplicial combinations generate better results than the
simple average of forecasts, the figures vary depending on the number of
forecasts in relation to the observations (years) used to construct the
weights. In table 4, we include the summary for each type of combination.
The SIMPLEX column compares varying-weight combinations ($S\_STC$\ plus $CAS$%
) with the fixed-weight one ($AVE$). When $J<T$, $CAS$ is 6.89 points higher
than $AVE$, but when $J>T$, this figure goes to 14.41 (more than twice
larger); that is, selection of forecasts works very well when its number is
greater than that of the years used to compute the weights, precisely when
some other methods can't say anything about it.\bigskip

\subsection{Results by method of combination, variable and accuracy criteria%
\protect\bigskip}

\begin{center}
\emph{Insert table 5 around here}\bigskip
\end{center}

Table 5 shows the percentage of beats by variable and accuracy criteria for
each combination procedure. The following comments are worth mentioning:

\begin{enumerate}
\item Fixed weights works better with RMSE\ and MAPE, although it never
reaches 50\% of cases and, on average, never beats $CAS$.

\item $CAS$ is the best with Mean Error and MdAPE, reaching, on average,
50\% of cases.

\item $S\_STC$ full combination is the worst except for $PGDP$ with MAPE\
and MdAPE, matching $CAS$ with Mean Error. This is also true for $RLSGOV$
with MdAPE.

\item $AVE$ is the best for $NGDP$, $EMP$, $TBILL$, $RRESIN$ and $RFEDGOV$
while $CAS$ is the best for $INDPROD$, $HOUSING$, $BOND$, $RGDP$, $RCONSUM$
and $RNRESIN$. Being the best is never achieved by the full $S\_STC$
combination.
\end{enumerate}

\subsection{Results by number of forecasts and accuracy criteria}

Table 6 shows the results of each combination by the number of forecasts and
accuracy criteria.\bigskip

\begin{center}
\emph{Insert table 6, panel a) and b) around here}\bigskip
\end{center}

In panel a), we present the number of beats of each combination which we
then added by fixed ($AVE$) and varying ($SIMPLEX$) weights in panel b).
There is a clear advantage of $SIMPLEX$ by around 30 points with Mean Error
and MdAPE. The difference is even greater when $J>T$. As we can see from
panel a), this is due to $CAS$. The difference further increases when $J>T$.
With $MdAPE$, for example, the difference goes from 7.19 percentage points
when $J<T$ to 25.53 when $J>T$ with respect to $AVE$.

\subsection{Results according to the variability of the forecasts}

The basic idea under this section is the following: a fixed-weight
combination assigns the same weight to forecasts, so if variability among
them is small, then the average will work well in the same direction,
however wrong it may be ('precisely' wrong) unless they are unbiased. On the
other hand, when variability is high, it is better to assign different
weights. This is in line with the results obtained by Jose and Winkler
(2008) when comparing the accuracy of the average with trimmed and
Winsorized averages and the results by Genre et al (2016) using the ECB
survey of professional forecasters. In this later paper, they find that some
combination methods outperform the simple average of forecasts in variables
with heterogeneity of forecasters and apparent bias.

In order to verify this hypothesis, we compute the variation coefficient
(VC)\ of each variable for each combination and forecast period from 2015 to
2018. We also plotted the forecasts for each period \footnote{%
In order to save space, these results are available upon request.}. In fact,
this issue forms part of the selection procedure presented in this paper.
The method is based on the orthogonality of forecasts; that is, it looks for
selecting those forecasts that do not share common information. In this
empirical application, the forecasts come from the Survey of Professional
Forecasters (SPF) and may have common information in forming their
forecasts. This is the reason why we expect some forecasts to be highly
correlated (even redundant) and others with low correlation. Then, the $CAS$
procedure takes advantage of this situation and usually generates better
results.

The main comments that can be pointed out are the following:

\begin{enumerate}
\item When all the forecasts included in the sample are highly correlated
and their plots show a similar behavior, $AVE$ is usually the best
combination. A clear example of this situation is shown in figure 4 where we
plot the forecasts for the variable $NGDP$ for all the samples.

\item When some of the forecasts are correlated but their plots differ
somewhat, $S\_STC$ is better because of its varying-weight allocation.
Figure 5 shows this situation for the variable $RLSGOV$.

\item In a mixed situation with some forecasts highly correlated and some
others not so, $CAS$\ is the best because it only selects non-redundant
forecasts. In figure 6 we show this situation for the variable $UNEMP$.

\item In general, with low correlated forecasts, varying-weight combinations
generate better results: the $S\_STC$ procedure, when the forecasts show a
similar behavior, and the $CAS$ procedure, when they don't. Figure 7 shows a
clear example of this situation for the variable $HOUSING$.
\end{enumerate}

Table 7 shows the VC and results for the aforementioned variables$\footnote{%
The VC, figures and results for the other variables are available upon
request. They have been omitted to save space.}$. The analysis of the VC
will be done jointly with figures 1 to 4.\bigskip 

\begin{center}
\emph{Insert table 7 and figures 1 to 4 arpund here}\bigskip 
\end{center}

\begin{enumerate}
\item \textbf{Variable NGDP}: All the graphs plotted in figure 1 show very
little variation between forecasts. The VC in each sample is very low,
suggesting that $AVE$ should be used. Looking at the combination results, $%
AVE$ is the winner in all the samples with the exception of sample 4. In
this case, $CAS$ generates the best forecasts for all the forecasting
periods. Notice that in the graph for sample 4, although the forecasts
follow a similar behavior, there are some of them with different patterns
that can be used to improve the forecast combination through the $CAS$
method.

\item \textbf{Variable RLSGOV}: The behavior of the forecasts for this
variable is different from the one observed before. In this case, the
forecasts seem to have a similar behavior, but the correlation between them
is not too high. Then, assigning different weights generates better
combinations. Looking at figure 2, we can see that $S\_STC$ obtains very
good results in 2017 and perhaps in 2016. Our perception from the graph is
confirmed in table 7: varying-weight combinations outperform the
fixed-weight one. This situation is also supported by the VC, which shows
higher values than the observed for the $NGDP$. So, in this case, the fact
that not all the forecasts show the same pattern leads to better forecasting
results with varying-weight methods.

\item \textbf{Variable UNEMP}: The VC of this variable in table 7 clearly
shows higher values than the observed for the previous variables. This fact
can indicate that the average forecast may not be the best combination in
this case. Looking at figure 3, not all the forecasts have the same pattern.
This favors the varying-weight combinations, $S\_STC$ and $CAS$, the latter
being the one that beats more times. Therefore, in this case, selection is
better than a full combination either fixed $AVE$ or varying $S\_STC$.

\item \textbf{Variable HOUSING}: Figure 4 shows the behavior of the
forecasts for the variable $HOUSING$. This is a clear example for $CAS$ to
form a combination. Different behavior of some forecasts and high VC are the
clues to select forecasts to obtain better forecasting results. Although
there is a common behavior of some forecasts, the selection of
orthogonalized forecasts improves the results.
\end{enumerate}

Similar results are confirmed for the other variables analyzed in the
empirical application. As a matter of fact, high VC and different behavior
might be the clues to consider CAS as the best subcombination to forecast a
variable.

\subsection{Results according to the forecast ability}

When the Diebold and Mariano (1995) or Giacomini and White (2006) tests are
not appropriate, it might be interesting to break down the MSFE into three
components (bias, variance, and covariance) to assess which of them holds
sway over a given MSFE:%
\begin{eqnarray*}
MSFE &:&=\frac{1}{H}\dsum\limits_{h=1}^{H}\left( \widehat{Y}%
_{T+h}-Y_{T+h}\right) ^{2}\equiv \left( \overline{\widehat{Y}}_{H}-\overline{%
Y}_{H}\right) ^{2}+\left( sd(\widehat{Y}_{f})-sd(Y_{f})\right) ^{2} \\
&&+2\left( sd(Y_{f})\right) \left( sd(Y_{f})\right) \left( 1-corr[\widehat{Y}%
_{f},Y_{f}]\right)
\end{eqnarray*}%
where $\overline{\widehat{Y}}_{H}$ is an $H$-period average forecast, $%
\overline{Y}_{H}$ is the corresponding average for the realized values, $sd(%
\widehat{Y}_{f})$ is the standard deviation of the forecasts, $sd(Y_{f})$ is
the standard deviation of the realized values, and $corr[\widehat{Y}%
_{f},Y_{f}]$ is the correlation between forecasts and realized values. Then,
proportions are defined as follow:

\[
\text{Bias proportion: }\frac{\left( \overline{\widehat{Y}}_{H}-\overline{Y}%
_{H}\right) ^{2}}{MSFE} 
\]

\[
\text{Variance proportion: }\frac{\left( sd(\widehat{Y}_{f})-sd(Y_{f})%
\right) ^{2}}{MSFE} 
\]

\[
\text{Covariance proportion: }\frac{2\left( sd(Y_{f})\right) \left(
sd(Y_{f})\right) \left( 1-corr[\widehat{Y}_{f},Y_{f}]\right) }{MSFE} 
\]%
Finally, we study which one constributes more to the MSFE. A ranking of
preferences may be given by the following four situations:

\begin{enumerate}
\item The best will be when there are little bias and variance (hence, high
covariance proportion).

\item The next one will be when there is little bias, but high variance
(hence, low covariance proportion).

\item Bad situations happen when the bias is high: either with high variance,

\item Or the worst, with low variance ('precisely' wrong).
\end{enumerate}

Using this classification, we show in Table 8 the bias, variance, and
covariance proportions for the combination procedures with lowest MSFE%
\footnote{%
The specific values for the bias, variance, and covariance proportions for
each variable, each sample, and each combination procedure are available
upon request.}.\bigskip

\begin{center}
Insert table 8 around here\bigskip
\end{center}

Looking at the results shown in Table 8, we can conclude that the $CAS$\
combination with lowest MSFE is classified in the best situation more than
50\% of the time, whereas the $AVE$ does in the third situation almost 50\%
of the time. The $S\_STC$ combination is also classified in the third
situation 60\% of the time. So, as a general result, we can conclude that $%
CAS$ shows the lowest MSFE where it is 'precisely' right.

\section{Conclusions}

In this paper, we have used the Split-Then-Combine ($STC$) approach to build
positive weights that sum up to one. Because of these two restrictions
(positiveness and adding up to 1), most methods from multivariate statistics
are inapplicable for combinational datasets, giving rise to a number of
issues that make inappropiate the Euclidean geometry. Instead, the Aitchison
geometry considers combinations of forecasts inside the simplex, the
sampling space of positive weights adding up to one. Basic transformations
from the simplex space to the real space and back to the simplex space allow
us to define different simplicial combinations with time-varying weights. In
addition, we develop new strategies to construct Combinations after
Selection (CAS) simplicial subcombinations by selecting those forecasts in a
full simplicial combination that assign higher weights than the one
allocated by the benchmark average or, where appropriate, a simplicial
subcombination based on orthogonal clusters of redundant forecasts.

The methodology can be summarized in these steps: first, we split experts'
forecasts by seasons to assess their relative forecast performance that
periodically evolve over time. Second, we choose as a combination vector the
gravity center of the simplex by means of an isometric, centered logratio
transformation. Then, we select forecasts inside a simplex of lower
dimension. Finally, we make rolling, truly out-of-sample, one-step-ahead
combinations of forecasts, even in cases where the sample size is smaller
than the number of forecasts. Once a new observation is known, we
recalculate the weights that we then keep one-step-ahead to form a new
out-of-sample combination.

We present experimental results with a pool of expert forecasters of the US
macroeconomy over the period 1991--2018. In most cases, the Combination
after Selection strategy improves the average (neutral combination in the
simplex space) with different criteria of forecasting accuracy, and works
very well even when the number of forecasts is greater than the number of
observations.

As a general rule, we can conclude that when there are a high number of
heterogeneous forecasts to be combined, the best way to form a combination
is by selecting a CAS simplicial subcombination formed by the most weighted,
non-redundant forecasts.\bigskip

\bigskip

Compliance with ethical standards:

Ethical approval: This article does not contain any studies with human or
animal participants performed by any of the authors.

\bigskip

Funding: Author A. de Juan Fern\'{a}ndez has received research grants from
Ministry of Economics of the Spanish government. Grant numbers:
ECO2015-70331-C2-1-R and PID2019-108079GB-C22.

Any of the authors have no conficts of interest

\newpage Tables: \bigskip

\begin{center}
Table 1: STC approach in the Euclidean Space

\hspace{-4cm}%
\begin{tabular}{c|cccc|cc}
Panel $m$ & 1 & 2 & ... & $J$ & $\overline{\widehat{Y}}_{J,t}^{(m)}$ & 
\multicolumn{1}{|c}{Real data} \\ \hline
1 & $\widehat{Y}_{1,1}^{(m)}$ & $\widehat{Y}_{2,1}^{(m)}$ & ... & $\widehat{Y%
}_{J,1}^{(m)}$ & $\overline{\widehat{Y}}_{J,1}^{(m)}$ & \multicolumn{1}{|c}{$%
Y_{1}^{(m)}$} \\ 
2 & $\widehat{Y}_{1,2}^{(m)}$ & $\widehat{Y}_{2,2}^{(m)}$ & ... & $\widehat{Y%
}_{J,2}^{(m)}$ & $\overline{\widehat{Y}}_{J,2}^{(m)}$ & \multicolumn{1}{|c}{$%
Y_{2}^{(m)}$} \\ 
... & ... & ... & ... & ... & ... & \multicolumn{1}{|c}{...} \\ 
$T_{1}$ & $\widehat{Y}_{1,T_{1}}^{(m)}$ & $\widehat{Y}_{2,T_{1}}^{(m)}$ & ...
& $\widehat{Y}_{J,T_{1}}^{(m)}$ & $\overline{\widehat{Y}}_{J,T_{1}}^{(m)}$ & 
\multicolumn{1}{|c}{$Y_{T_{1}}^{(m)}$} \\ \hline
$\overline{\widehat{Y}}_{j,T_{1}}$ & $\overline{\widehat{Y}}_{1,T_{1}}$ & $%
\overline{\widehat{Y}}_{2,T_{1}}$ & ... & $\overline{\widehat{Y}}_{J,T_{1}}$
& $\overline{\overline{\widehat{Y}}}_{J,T_{1}}^{(m)}$ & \multicolumn{1}{|c}{$%
\overline{Y}_{T_{1}}^{(m)}$} \\ \hline
Fixed & $\left( \overline{\widehat{Y}}_{1,T_{1}}^{(m)}-\overline{\overline{%
\widehat{Y}}}_{J,T_{1}}^{(m)}\right) ^{2}$ & $\left( \overline{\widehat{Y}}%
_{2,T_{1}}^{(m)}-\overline{\overline{\widehat{Y}}}_{J,T_{1}}^{(m)}\right)
^{2}$ & ... & $\left( \overline{\widehat{Y}}_{J,T_{1}}^{(m)}-\overline{%
\overline{\widehat{Y}}}_{J,T_{1}}^{(m)}\right) ^{2}$ &  &  \\ \cline{1-5}
\end{tabular}
\end{center}

\newpage

\bigskip

\begin{center}
Table 2: Definition of the main variables used in the application.

\hspace{-4cm}%
\begin{tabular}{ccc}
{\small Variable} & {\small Definition} & {\small Sample} \\ \hline
\multicolumn{1}{l}{${\small NGDP}$} & \multicolumn{1}{l}{\small Forecasts
for the quarterly level of nominal GDP. SA. billions \$} & \multicolumn{1}{l}%
{\small 1991 Q1 - 2018 Q4} \\ \hline
\multicolumn{1}{l}{${\small PGDP}$} & \multicolumn{1}{l}{\small Forecasts
for the quaterly level of the chian-weighted GDP price index.} & 
\multicolumn{1}{l}{} \\ 
\multicolumn{1}{l}{} & \multicolumn{1}{l}{\small SA. Index. Base year 1992}
& \multicolumn{1}{l}{\small 1991 Q1 - 2018 Q4} \\ \hline
\multicolumn{1}{l}{${\small UNEMP}$} & \multicolumn{1}{l}{\small Forecasts
for the quarterly average unemployment rate. SA. \% points} & 
\multicolumn{1}{l}{\small 1991 Q1 - 2018 Q4} \\ \hline
\multicolumn{1}{l}{${\small EMP}$} & \multicolumn{1}{l}{\small Forecasts for
the quarterly average level of nonfarm payroll employment.} & 
\multicolumn{1}{l}{} \\ 
\multicolumn{1}{l}{} & \multicolumn{1}{l}{\small SA. Thousands of jobs.} & 
\multicolumn{1}{l}{\small 2004 Q1 - 2018 Q4} \\ \hline
\multicolumn{1}{l}{${\small INDPROD}$} & \multicolumn{1}{l}{\small Forecasts
for the quarterly average level of the index of industrial prod.} & 
\multicolumn{1}{l}{} \\ 
\multicolumn{1}{l}{} & \multicolumn{1}{l}{\small SA. Index.} & 
\multicolumn{1}{l}{\small 1991 Q1 - 2018 Q4} \\ \hline
\multicolumn{1}{l}{${\small HOUSING}$} & \multicolumn{1}{l}{\small Forecasts
for the quarterly average level of housing starts. SA. millions.} & 
\multicolumn{1}{l}{\small 1991 Q1 - 2018 Q4} \\ \hline
\multicolumn{1}{l}{${\small TBILL}$} & \multicolumn{1}{l}{\small Forecasts
for the quarterly average 3-months Treasury Bill rates. \% points} & 
\multicolumn{1}{l}{\small 1991 Q1 - 2018 Q4} \\ \hline
\multicolumn{1}{l}{${\small BOND}$} & \multicolumn{1}{l}{\small Forecasts
for the quarterly average level of Moody's Aaa corporate} & 
\multicolumn{1}{l}{} \\ 
\multicolumn{1}{l}{} & \multicolumn{1}{l}{\small Bond yield. \% points} & 
\multicolumn{1}{l}{\small 1991 Q1 - 2018 Q4} \\ \hline
\multicolumn{1}{l}{${\small RGDP}$} & \multicolumn{1}{l}{\small Forecasts
for the quarterly chain-weighted real GDP. SA. annual rate.} & 
\multicolumn{1}{l}{} \\ 
\multicolumn{1}{l}{} & \multicolumn{1}{l}{\small Base years 1992 - 1995,
fixed weighted real GDP} & \multicolumn{1}{l}{\small 1991 Q1 - 2018 Q4} \\ 
\hline
\multicolumn{1}{l}{${\small RCONSUM}$} & \multicolumn{1}{l}{\small Forecasts
for the quarterly chain-weighted real personal consumption} & 
\multicolumn{1}{l}{} \\ 
\multicolumn{1}{l}{} & \multicolumn{1}{l}{\small expenditures. SA, annual
rate, \ base years 1992 - 1995.} & \multicolumn{1}{l}{\small 1991 Q1 - 2018
Q4} \\ \hline
\multicolumn{1}{l}{${\small RNRESIN}$} & \multicolumn{1}{l}{\small Forecasts
for the quarterly chain-weighted real nonresidential fixed} & 
\multicolumn{1}{l}{} \\ 
\multicolumn{1}{l}{} & \multicolumn{1}{l}{\small investment. SA. annual
rate, base years 1992 - 1995.} & \multicolumn{1}{l}{\small 1991 Q1 - 2018 Q4}
\\ \hline
\multicolumn{1}{l}{${\small RRESINV}$} & \multicolumn{1}{l}{\small Forecasts
for the quarterly chain-weighted real residential fixed} & 
\multicolumn{1}{l}{} \\ 
\multicolumn{1}{l}{} & \multicolumn{1}{l}{\small investment. SA., annual
rate, base years 1992 - 1995} & \multicolumn{1}{l}{\small 1991 Q1 - 2018 Q4}
\\ \hline
\multicolumn{1}{l}{${\small RFEDGOV}$} & \multicolumn{1}{l}{\small Forecasts
for the quarterly chain-weighted real federal government} & 
\multicolumn{1}{l}{} \\ 
\multicolumn{1}{l}{} & \multicolumn{1}{l}{\small consumption and gross
investment. SA, annual rate, base years 1992-95} & \multicolumn{1}{l}{\small %
1991 Q1 - 2018 Q4} \\ \hline
\multicolumn{1}{l}{${\small RLSGOV}$} & \multicolumn{1}{l}{\small Forecasts
for the quarterly level of chain-weighted real state and local} & 
\multicolumn{1}{l}{} \\ 
\multicolumn{1}{l}{} & \multicolumn{1}{l}{\small government consumption and
gross investment. SA. annual rate.} & \multicolumn{1}{l}{} \\ 
\multicolumn{1}{l}{} & \multicolumn{1}{l}{\small base years 1992 - 1995} & 
\multicolumn{1}{l}{\small 1991 Q1 - 2018 Q4} \\ \hline
\multicolumn{1}{l}{${\small CPI}$} & \multicolumn{1}{l}{\small Forecasts for
the headline CPI inflation rate. SA, annual rate, \% points.} & 
\multicolumn{1}{l}{} \\ 
\multicolumn{1}{l}{} & \multicolumn{1}{l}{\small Quarterly forecasts are
annualized quarter-overquarter percent changes} & \multicolumn{1}{l}{} \\ 
\multicolumn{1}{l}{} & \multicolumn{1}{l}{\small of the quarterly average
price index level} & \multicolumn{1}{l}{\small 1991 Q1 - 2018 Q4} \\ \hline
\end{tabular}

Source: Survey of Professional Forecasters documentation. SA = Seasonal
Adjusted.

\newpage

\bigskip

Table 3: Variables, samples and number of forecasters

\begin{tabular}{l|cc|c|c|cc|c|c|cc}
& \multicolumn{10}{c}{${\small Samples}$} \\ \cline{2-11}
${\small Variable}$ & \multicolumn{2}{|l|}{${\small Sample\ (1)}$} & 
\multicolumn{2}{|l|}{${\small Sample\ (2)}$} & \multicolumn{2}{|l|}{${\small %
Sample\ (3)}$} & \multicolumn{2}{|l|}{${\small Sample\ (4)}$} & 
\multicolumn{2}{|l}{${\small Sample\ (5)}$} \\ \hline
& ${\small T}$ & ${\small J}$ & ${\small T}$ & ${\small J}$ & ${\small T}$ & 
${\small J}$ & ${\small T}$ & ${\small J}$ & ${\small T}$ & ${\small J}$ \\ 
\cline{2-11}
${\small NGDP}$ & ${\small 24}$ & ${\small 3}$ & ${\small 20}^{a)}$ & $%
{\small 6}$ & ${\small 15}^{d)}$ & ${\small 10}$ & ${\small 9}^{g)}$ & $%
{\small 18}$ & ${\small 5}$ & ${\small 22}$ \\ 
${\small PGDP}$ & ${\small 24}$ & ${\small 3}$ & ${\small 20}^{a)}$ & $%
{\small 6}$ & ${\small 15}^{d)}$ & ${\small 10}$ & ${\small 9}^{g)}$ & $%
{\small 20}$ & ${\small 5}$ & ${\small 25}$ \\ 
${\small UNEMP}$ & ${\small 24}$ & ${\small 4}$ & ${\small 20}^{a)}$ & $%
{\small 6}$ & ${\small 15}^{d)}$ & ${\small 12}$ & ${\small 9}^{g)}$ & $%
{\small 22}$ & ${\small 5}$ & ${\small 27}$ \\ 
${\small EMP}$ &  &  & ${\small 11}$ & ${\small 16}$ & ${\small 10}$ & $%
{\small 20}$ & ${\small 8}$ & ${\small 22}$ & ${\small 5}$ & ${\small 28}$
\\ 
${\small INDPROD}$ & ${\small 24}$ & ${\small 4}$ & ${\small 19}^{b)}$ & $%
{\small 8}$ & ${\small 15}^{d)}$ & ${\small 12}$ & ${\small 9}^{g)}$ & $%
{\small 21}$ & ${\small 5}$ & ${\small 26}$ \\ 
${\small HOUSING}$ & ${\small 24}$ & ${\small 4}$ & ${\small 19}^{b)}$ & $%
{\small 10}$ & ${\small 15}^{d)}$ & ${\small 15}$ & ${\small 10}^{f)}$ & $%
{\small 19}$ & ${\small 5}$ & ${\small 26}$ \\ 
${\small TBILL}$ & ${\small 24}$ & ${\small 5}$ & ${\small 19}^{b)}$ & $%
{\small 8}$ & ${\small 15}^{d)}$ & ${\small 11}$ & ${\small 9}^{g)}$ & $%
{\small 19}$ & ${\small 5}$ & ${\small 24}$ \\ 
${\small BOND}$ & ${\small 24}$ & ${\small 3}$ & ${\small 19}^{b)}$ & $%
{\small 5}$ & ${\small 14}^{e)}$ & ${\small 7}$ & ${\small 9}^{g)}$ & $%
{\small 13}$ & ${\small 5}$ & ${\small 17}$ \\ 
${\small RRESINV}$ & ${\small 24}$ & ${\small 5}$ & ${\small 20}^{a)}$ & $%
{\small 9}$ & ${\small 15}^{d)}$ & ${\small 13}$ & ${\small 9}^{g)}$ & $%
{\small 19}$ & ${\small 5}$ & ${\small 28}$ \\ 
${\small RGDP}$ & ${\small 24}$ & ${\small 5}$ & ${\small 20}^{a)}$ & $%
{\small 9}$ & ${\small 15}^{d)}$ & ${\small 14}$ & ${\small 9}^{g)}$ & $%
{\small 25}$ & ${\small 5}$ & ${\small 31}$ \\ 
${\small RCONSUM}$ & ${\small 24}$ & ${\small 5}$ & ${\small 20}^{a)}$ & $%
{\small 9}$ & ${\small 16}^{c)}$ & ${\small 13}$ & ${\small 10}^{f)}$ & $%
{\small 20}$ & ${\small 5}$ & ${\small 29}$ \\ 
${\small RNREIN}$ & ${\small 24}$ & ${\small 5}$ & ${\small 20}^{a)}$ & $%
{\small 9}$ & ${\small 16}^{c)}$ & ${\small 13}$ & ${\small 10}^{f)}$ & $%
{\small 20}$ & ${\small 5}$ & ${\small 29}$ \\ 
${\small RFEDGOV}$ & ${\small 24}$ & ${\small 5}$ & ${\small 20}^{a)}$ & $%
{\small 9}$ & ${\small 16}^{c)}$ & ${\small 13}$ & ${\small 10}^{f)}$ & $%
{\small 19}$ & ${\small 5}$ & ${\small 28}$ \\ 
${\small RLSGOV}$ & ${\small 24}$ & ${\small 5}$ & ${\small 20}^{a)}$ & $%
{\small 9}$ & ${\small 16}^{c)}$ & ${\small 13}$ & ${\small 10}^{f)}$ & $%
{\small 19}$ & ${\small 5}$ & ${\small 28}$ \\ 
${\small CPI}$ & ${\small 24}$ & ${\small 5}$ & ${\small 20}^{a)}$ & $%
{\small 8}$ & ${\small 16}^{c)}$ & ${\small 12}$ & ${\small 10}^{f)}$ & $%
{\small 19}$ & ${\small 5}$ & ${\small 29}$ \\ \hline
\end{tabular}

${\small T=}$ {\small number of periods, }${\small J=}$ {\small number of
forecasters, Sample (1): 1991 - 2014; Sample (2) a) 1995 - 2014; b) 1996 -
2014; Sample (3) c) 1999 - 2014; d) 2000 - 2014; e) 2001 - 2014; Sample (4)
f) 2005 - 2014; g) 2006 - 2014; Sample (5) 2010 - 2014; For the EMP variable
the samples are: (1) 2004-2014; (2) 2005-2014; (3) 2007-2014 and (4)
2010-2014}

\bigskip

Table 4: Summary between combinations depending on $J$ and $T$

\begin{tabular}{lccc|c|c}
& $AVERAGE$ & $S\_STC$ & $CAS$ & $SIMPLEX$ & $TOTAL$ \\ \cline{2-6}
$J<T$ & $282$ & $111$ & $332$ & $443$ & $725$ \\ 
$(\%)$ & $(38.90)$ & $(15.31)$ & $(45.79)$ & $(61.10)$ & $(57.27)$ \\ \hline
$J>T$ & $189$ & $85$ & $267$ & $352$ & $541$ \\ 
$(\%)$ & $(34.94)$ & $(15.71)$ & $(49.35)$ & $(65.06)$ & $(42.73)$ \\ \hline
$TOTAL$ & $469$ & $194$ & $603$ & $797$ & $1266$ \\ 
$(\%)$ & $(37.05)$ & $(15.32)$ & $(47.63)$ & $(62.95)$ &  \\ \cline{1-5}
\end{tabular}

{\small Number of times that an accuracy measure favored a combination
procedure.}

\newpage
\end{center}

\landscape%

\begin{center}
Table 5: Results for each combination procedure by variable and accuracy
criteria. Percentages of beats

\begin{tabular}{l|ccc|ccc|ccc|ccc|}
& \multicolumn{3}{|c|}{\small Mean Error} & \multicolumn{3}{|c|}{\small RMSE}
& \multicolumn{3}{|c|}{\small MAPE} & \multicolumn{3}{|c|}{\small MdAPE} \\ 
\cline{2-13}
& ${\small AVE}$ & ${\small S\_STC}$ & ${\small CAS}$ & ${\small AVE}$ & $%
{\small S\_\ STC}$ & ${\small CAS}$ & ${\small AVE}$ & ${\small S\_STC}$ & $%
{\small CAS}$ & ${\small AVE}$ & ${\small S\_STC}$ & ${\small CAS}$ \\ \hline
\multicolumn{1}{|c|}{${\small NGDP}$} & ${\small 47.62}$ & ${\small 14.29}$
& ${\small 38.10}$ & ${\small 47.83}$ & ${\small 13.04}$ & ${\small 39.13}$
& ${\small 56.52}$ & ${\small 13.04}$ & ${\small 30.43}$ & ${\small 66.67}$
& ${\small 9.52}$ & ${\small 23.81}$ \\ 
\multicolumn{1}{|c|}{${\small PGDP}$} & ${\small 21.74}$ & ${\small 39.13}$
& ${\small 39.13}$ & ${\small 41.67}$ & ${\small 33.33}$ & ${\small 25.00}$
& ${\small 30.77}$ & ${\small 42.31}$ & ${\small 26.92}$ & ${\small 32.00}$
& ${\small 40.00}$ & ${\small 28.00}$ \\ 
\multicolumn{1}{|c|}{${\small UNEMP}$} & ${\small 28.57}$ & ${\small 9.52}$
& ${\small 61.9}$ & ${\small 42.86}$ & ${\small 14.29}$ & ${\small 42.86}$ & 
${\small 38.10}$ & ${\small 9.52}$ & ${\small 52.38}$ & ${\small 38.10}$ & $%
{\small 9.52}$ & ${\small 52.38}$ \\ 
\multicolumn{1}{|c|}{${\small EMP}$} & ${\small 72.22}$ & ${\small 11.11}$ & 
${\small 16.67}$ & ${\small 82.35}$ & ${\small 5.88}$ & ${\small 11.76}$ & $%
{\small 78.95}$ & ${\small 5.26}$ & ${\small 15.79}$ & ${\small 78.95}$ & $%
{\small 5.26}$ & ${\small 15.79}$ \\ 
\multicolumn{1}{|c|}{${\small INDPROD}$} & ${\small 20.00}$ & ${\small 25.00}
$ & ${\small 55.00}$ & ${\small 25.00}$ & ${\small 5.00}$ & ${\small 70.00}$
& ${\small 23.81}$ & ${\small 9.52}$ & ${\small 66.67}$ & ${\small 9.09}$ & $%
{\small 13.64}$ & ${\small 77.27}$ \\ 
\multicolumn{1}{|c|}{${\small HOUSING}$} & ${\small 0.00}$ & ${\small 15.00}$
& ${\small 85.00}$ & ${\small 4.76}$ & ${\small 19.05}$ & ${\small 76.19}$ & 
${\small 4.76}$ & ${\small 9.52}$ & ${\small 85.71}$ & ${\small 0.00}$ & $%
{\small 9.09}$ & ${\small 90.91}$ \\ 
\multicolumn{1}{|c|}{${\small TBILL}$} & ${\small 75.00}$ & ${\small 0.00}$
& ${\small 25.00}$ & ${\small 75.00}$ & ${\small 5.00}$ & ${\small 20.00}$ & 
${\small 70.00}$ & ${\small 10.00}$ & ${\small 20.00}$ & ${\small 50.00}$ & $%
{\small 9.09}$ & ${\small 40.91}$ \\ 
\multicolumn{1}{|c|}{${\small BOND}$} & ${\small 10.00}$ & ${\small 5.00}$ & 
${\small 85.00}$ & ${\small 20.00}$ & ${\small 10.00}$ & ${\small 70.00}$ & $%
{\small 15.00}$ & ${\small 15.00}$ & ${\small 70.00}$ & ${\small 10.00}$ & $%
{\small 15.00}$ & ${\small 75.00}$ \\ 
\multicolumn{1}{|c|}{${\small RRESIN}$} & ${\small 57.14}$ & ${\small 4.76}$
& ${\small 38.10}$ & ${\small 50.00}$ & ${\small 22.73}$ & ${\small 27.27}$
& ${\small 57.14}$ & ${\small 9.52}$ & ${\small 33.33}$ & ${\small 54.55}$ & 
${\small 9.09}$ & ${\small 36.36}$ \\ 
\multicolumn{1}{|c|}{${\small RGDP}$} & ${\small 4.76}$ & ${\small 4.76}$ & $%
{\small 90.48}$ & ${\small 8.70}$ & ${\small 17.39}$ & ${\small 73.91}$ & $%
{\small 9.09}$ & ${\small 13.64}$ & ${\small 77.27}$ & ${\small 9.09}$ & $%
{\small 9.09}$ & ${\small 81.82}$ \\ 
\multicolumn{1}{|c|}{${\small RCONSUM}$} & ${\small 13.64}$ & ${\small 18.18}
$ & ${\small 68.18}$ & ${\small 28.57}$ & ${\small 9.52}$ & ${\small 61.90}$
& ${\small 31.82}$ & ${\small 9.09}$ & ${\small 59.09}$ & ${\small 25.00}$ & 
${\small 5.00}$ & ${\small 70.00}$ \\ 
\multicolumn{1}{|c|}{${\small RNRESIN}$} & ${\small 20.00}$ & ${\small 25.00}
$ & ${\small 55.00}$ & ${\small 31.82}$ & ${\small 22.73}$ & ${\small 45.45}$
& ${\small 30.00}$ & ${\small 20.00}$ & ${\small 50.00}$ & ${\small 25.00}$
& ${\small 8.33}$ & ${\small 66.67}$ \\ 
\multicolumn{1}{|c|}{${\small RFEDGOV}$} & ${\small 42.86}$ & ${\small 19.05}
$ & ${\small 38.10}$ & ${\small 65.00}$ & ${\small 25.00}$ & ${\small 10.00}$
& ${\small 71.43}$ & ${\small 14.29}$ & ${\small 14.29}$ & ${\small 55.00}$
& ${\small 20.00}$ & ${\small 25.00}$ \\ 
\multicolumn{1}{|c|}{${\small RLSGOV}$} & ${\small 45.00}$ & ${\small 30.00}$
& ${\small 25.00}$ & ${\small 55.00}$ & ${\small 25.00}$ & ${\small 20.00}$
& ${\small 40.00}$ & ${\small 25.00}$ & ${\small 35.00}$ & ${\small 31.82}$
& ${\small 40.91}$ & ${\small 27.27}$ \\ 
\multicolumn{1}{|c|}{${\small CPI}$} & ${\small 52.63}$ & ${\small 10.53}$ & 
${\small 36.84}$ & ${\small 45.00}$ & ${\small 5.00}$ & ${\small 50.00}$ & $%
{\small 50.00}$ & ${\small 15.00}$ & ${\small 35.00}$ & ${\small 50.00}$ & $%
{\small 10.00}$ & ${\small 40.00}$ \\ \hline
\multicolumn{1}{|c|}{${\small MEAN}$} & ${\small 34.08}$ & ${\small 15.42}$
& ${\small 50.50}$ & ${\small 41.57}$ & ${\small 15.53}$ & ${\small 42.90}$
& ${\small 40.49}$ & ${\small 14.71}$ & ${\small 44.79}$ & ${\small 35.68}$
& ${\small 14.24}$ & ${\small 50.08}$ \\ \hline
\end{tabular}

\bigskip

\newpage
\end{center}

\bigskip

\begin{center}
Table 6 a) Number of beats of each combination by accuracy criteria and
number of forecasts.

\hspace{-3cm}%
\begin{tabular}{c|ccc|ccc|ccc|ccc|}
& \multicolumn{3}{|c|}{$Mean\ Error$} & \multicolumn{3}{|c|}{$RMSE$} & 
\multicolumn{3}{|c|}{$MAPE$} & \multicolumn{3}{|c|}{$MdAPE$} \\ 
\cline{2-13}\cline{8-10}
& $AVE$ & $S\_STC$ & $CAS$ & $AVE$ & $S\_STC$ & $CAS$ & $AVE$ & $S\_STC$ & $%
CAS$ & $AVE$ & $S\_STC$ & $CAS$ \\ \hline
$J<T$ & $58$ & $26$ & $92$ & $77$ & $31$ & $77$ & $77$ & $26$ & $80$ & $70$
& $28$ & $83$ \\ 
$(\%)$ & $(32.95)$ & $(14.77)$ & $(52.27)$ & $(41.62)$ & $(16.76)$ & $%
(41.62) $ & $(42.08)$ & $(14.21)$ & $(43.72)$ & $(38.67)$ & $(15.47)$ & $%
(45.86)$ \\ \hline
$J>T$ & $45$ & $22$ & $64$ & $51$ & $19$ & $62$ & $50$ & $25$ & $62$ & $43$
& $19$ & $79$ \\ 
$(\%)$ & $(34.35)$ & $(16.79)$ & $(48.85)$ & $(38.64)$ & $(14.39)$ & $%
(46.97) $ & $(36.50)$ & $(18.25)$ & $(45.26)$ & $(30.50)$ & $(13.48)$ & $%
(56.03)$ \\ \hline
$TOTAL$ & $103$ & $48$ & $156$ & $128$ & $50$ & $139$ & $127$ & $51$ & $142$
& $113$ & $47$ & $162$ \\ 
$(\%)$ & $(33.55)$ & $(15.64)$ & $(50.81)$ & $(40.38)$ & $(15.77)$ & $%
(43.85) $ & $(39.69)$ & $(15.94)$ & $(44.38)$ & $(35.09)$ & $(14.60)$ & $%
(50.31)$ \\ \hline
\end{tabular}

\bigskip

Table 6 b) Number of beats of AVERAGE and SIMPLEX combinations by accuracy
criteria and number of forecasts

\begin{tabular}{l|cc|cc|cc|cc|}
& \multicolumn{2}{|c|}{$Mean\ Error$} & \multicolumn{2}{|c|}{$RMSE$} & 
\multicolumn{2}{|c|}{$MAPE$} & \multicolumn{2}{|c|}{$MdAPE$} \\ \cline{2-9}
& $AVE$ & $SIMPLEX$ & $AVE$ & $SIMPLEX$ & $AVE$ & $SIMPLEX$ & $AVE$ & $%
SIMPLEX$ \\ \hline
$J<T$ & $58$ & $118$ & $77$ & $108$ & $77$ & $106$ & $70$ & $111$ \\ 
$(\%)$ & $(32.95)$ & $(67.05)$ & $(41.62)$ & $(58.38)$ & $(42.08)$ & $%
(57.92) $ & $(38.67)$ & $(61.33)$ \\ \hline
$J>T$ & $45$ & $86$ & $51$ & $81$ & $50$ & $87$ & $43$ & $98$ \\ 
$(\%)$ & $(34.35)$ & $(65.65)$ & $(38.64)$ & $(61.36)$ & $(36.50)$ & $%
(63.50) $ & $(30.50)$ & $(69.50)$ \\ \hline
$TOTAL$ & $103$ & $204$ & $128$ & $189$ & $127$ & $193$ & $113$ & $209$ \\ 
$(\%)$ & $(33.55)$ & $(66.45)$ & $(40.38)$ & $(59.62)$ & $(39.69)$ & $%
(60.31) $ & $(35.09)$ & $(64.91)$ \\ \hline
\end{tabular}

\bigskip

\newpage
\end{center}

\bigskip

\begin{center}
Table 7: Coefficients of variation for selected variables and results of the
combination procedures by samples

\hspace{-4cm}%
\begin{tabular}{ccccc|c|c|c|c|cccc|c|c|c|c}
& \multicolumn{4}{c|}{${\small NGDP}$} & \multicolumn{4}{|c|}{${\small UNEMP}
$} & \multicolumn{4}{|c|}{${\small RLSGOV}$} & \multicolumn{4}{|c}{${\small %
HOUSING}$} \\ \cline{2-17}
& ${\small CV}$ & ${\small AVE}$ & ${\small S\_STC}$ & ${\small CAS}$ & $%
{\small CV}$ & ${\small AVE}$ & ${\small S\_STC}$ & ${\small CAS}$ & $%
{\small CV}$ & ${\small AVE}$ & ${\small S\_STC}$ & ${\small CAS}$ & $%
{\small CV}$ & ${\small AVE}$ & ${\small S\_STC}$ & ${\small CAS}$ \\ \hline
{\small Sample (1)} &  & ${\small 12}$ & ${\small 4}$ & ${\small 0}$ &  & $%
{\small 6}$ & ${\small 3}$ & ${\small 7}$ &  & ${\small 9}$ & ${\small 6}$ & 
${\small 2}$ &  & ${\small 0}$ & ${\small 0}$ & ${\small 16}$ \\ \hline
{\small 2015} & ${\small 0.667}$ & ${\small 4}$ & ${\small 0}$ & ${\small 0}$
& ${\small 0.603}$ & ${\small 4}$ & ${\small 0}$ & ${\small 0}$ & ${\small %
0.588}$ & ${\small 3}$ & ${\small 1}$ & ${\small 0}$ & ${\small 7.970}$ & $%
{\small 0}$ & ${\small 0}$ & ${\small 4}$ \\ 
{\small 2016} & ${\small 0.405}$ & ${\small 0}$ & ${\small 4}$ & ${\small 0}$
& ${\small 2.150}$ & ${\small 0}$ & ${\small 3}$ & ${\small 1}$ & ${\small %
0.724}$ & ${\small 2}$ & ${\small 0}$ & ${\small 2}$ & ${\small 11.287}$ & $%
{\small 0}$ & ${\small 0}$ & ${\small 4}$ \\ 
{\small 2017} & ${\small 0.313}$ & ${\small 4}$ & ${\small 0}$ & ${\small 0}$
& ${\small 4.927}$ & ${\small 0}$ & ${\small 0}$ & ${\small 4}$ & ${\small %
0.695}$ & ${\small 0}$ & ${\small 4}$ & ${\small 0}$ & ${\small 9.604}$ & $%
{\small 0}$ & ${\small 0}$ & ${\small 4}$ \\ 
{\small 2018} & ${\small 0.409}$ & ${\small 4}$ & ${\small 0}$ & ${\small 0}$
& ${\small 1.575}$ & ${\small 2}$ & ${\small 0}$ & ${\small 2}$ & ${\small %
0.324}$ & ${\small 4}$ & ${\small 1}$ & ${\small 0}$ & ${\small 9.850}$ & $%
{\small 0}$ & ${\small 4}$ & ${\small 4}$ \\ \hline
{\small Sample (2)} &  & ${\small 10}$ & ${\small 2}$ & ${\small 5}$ &  & $%
{\small 7}$ & ${\small 1}$ & ${\small 8}$ &  & ${\small 4}$ & ${\small 4}$ & 
${\small 8}$ &  & ${\small 2}$ & ${\small 0}$ & ${\small 10}$ \\ \hline
{\small 2015} & ${\small 0.830}$ & ${\small 4}$ & ${\small 0}$ & ${\small 0}$
& ${\small 3.447}$ & ${\small 1}$ & ${\small 0}$ & ${\small 3}$ & ${\small %
0.834}$ & ${\small 0}$ & ${\small 0}$ & ${\small 4}$ & ${\small 6.813}$ & $%
{\small 0}$ & ${\small 1}$ & ${\small 4}$ \\ 
{\small 2016} & ${\small 0.365}$ & ${\small 2}$ & ${\small 1}$ & ${\small 2}$
& ${\small 2.011}$ & ${\small 3}$ & ${\small 1}$ & ${\small 0}$ & ${\small %
0.627}$ & ${\small 3}$ & ${\small 1}$ & ${\small 0}$ & ${\small 9.079}$ & $%
{\small 0}$ & ${\small 2}$ & ${\small 3}$ \\ 
{\small 2017} & ${\small 0.245}$ & ${\small 0}$ & ${\small 1}$ & ${\small 3}$
& ${\small 4.080}$ & ${\small 0}$ & ${\small 0}$ & ${\small 4}$ & ${\small %
0.566}$ & ${\small 1}$ & ${\small 3}$ & ${\small 0}$ & ${\small 8.518}$ & $%
{\small 0}$ & ${\small 1}$ & ${\small 2}$ \\ 
{\small 2018} & ${\small 0.465}$ & ${\small 4}$ & ${\small 0}$ & ${\small 0}$
& ${\small 2.434}$ & ${\small 3}$ & ${\small 0}$ & ${\small 1}$ & ${\small %
1.666}$ & ${\small 0}$ & ${\small 0}$ & ${\small 4}$ & ${\small 7.486}$ & $%
{\small 2}$ & ${\small 0}$ & ${\small 1}$ \\ \hline
{\small Sample (3)} &  & ${\small 11}$ & ${\small 5}$ & ${\small 4}$ &  & $%
{\small 7}$ & ${\small 2}$ & ${\small 8}$ &  & ${\small 6}$ & ${\small 4}$ & 
${\small 6}$ &  & ${\small 0}$ & ${\small 0}$ & ${\small 16}$ \\ \hline
{\small 2015} & ${\small 0.741}$ & ${\small 4}$ & ${\small 0}$ & ${\small 0}$
& ${\small 2.743}$ & ${\small 1}$ & ${\small 1}$ & ${\small 3}$ & ${\small %
0.807}$ & ${\small 2}$ & ${\small 0}$ & ${\small 2}$ & ${\small 6.575}$ & $%
{\small 0}$ & ${\small 0}$ & ${\small 4}$ \\ 
{\small 2016} & ${\small 0.342}$ & ${\small 1}$ & ${\small 2}$ & ${\small 2}$
& ${\small 1.829}$ & ${\small 4}$ & ${\small 0}$ & ${\small 0}$ & ${\small %
0.550}$ & ${\small 3}$ & ${\small 1}$ & ${\small 0}$ & ${\small 8.528}$ & $%
{\small 0}$ & ${\small 0}$ & ${\small 4}$ \\ 
{\small 2017} & ${\small 0.217}$ & ${\small 2}$ & ${\small 3}$ & ${\small 2}$
& ${\small 3.804}$ & ${\small 0}$ & ${\small 0}$ & ${\small 4}$ & ${\small %
0.504}$ & ${\small 0}$ & ${\small 0}$ & ${\small 4}$ & ${\small 7.485}$ & $%
{\small 0}$ & ${\small 0}$ & ${\small 4}$ \\ 
{\small 2018} & ${\small 0.392}$ & ${\small 4}$ & ${\small 0}$ & ${\small 0}$
& ${\small 2.951}$ & ${\small 2}$ & ${\small 1}$ & ${\small 1}$ & ${\small %
1.418}$ & ${\small 1}$ & ${\small 3}$ & ${\small 0}$ & ${\small 6.476}$ & $%
{\small 0}$ & ${\small 0}$ & ${\small 4}$ \\ \hline
{\small Sample (4)} &  & ${\small 2}$ & ${\small 0}$ & ${\small 14}$ &  & $%
{\small 8}$ & ${\small 3}$ & ${\small 8}$ &  & ${\small 8}$ & ${\small 7}$ & 
${\small 2}$ &  & ${\small 0}$ & ${\small 2}$ & ${\small 15}$ \\ \hline
{\small 2015} & ${\small 0.590}$ & ${\small 1}$ & ${\small 0}$ & ${\small 3}$
& ${\small 2.253}$ & ${\small 0}$ & ${\small 0}$ & ${\small 4}$ & ${\small %
0.725}$ & ${\small 3}$ & ${\small 1}$ & ${\small 0}$ & ${\small 6.188}$ & $%
{\small 0}$ & ${\small 0}$ & ${\small 4}$ \\ 
{\small 2016} & ${\small 0.306}$ & ${\small 1}$ & ${\small 0}$ & ${\small 3}$
& ${\small 1.766}$ & ${\small 3}$ & ${\small 3}$ & ${\small 1}$ & ${\small %
0.499}$ & ${\small 1}$ & ${\small 1}$ & ${\small 2}$ & ${\small 7.665}$ & $%
{\small 0}$ & ${\small 1}$ & ${\small 3}$ \\ 
{\small 2017} & ${\small 0.248}$ & ${\small 0}$ & ${\small 0}$ & ${\small 4}$
& ${\small 3.104}$ & ${\small 1}$ & ${\small 0}$ & ${\small 3}$ & ${\small %
0.446}$ & ${\small 0}$ & ${\small 4}$ & ${\small 0}$ & ${\small 6.814}$ & $%
{\small 0}$ & ${\small 0}$ & ${\small 5}$ \\ 
{\small 2018} & ${\small 0.502}$ & ${\small 0}$ & ${\small 0}$ & ${\small 4}$
& ${\small 3.132}$ & ${\small 4}$ & ${\small 0}$ & ${\small 0}$ & ${\small %
1.751}$ & ${\small 4}$ & ${\small 1}$ & ${\small 0}$ & ${\small 6.247}$ & $%
{\small 0}$ & ${\small 1}$ & ${\small 3}$ \\ \hline
{\small Sample (5)} &  & ${\small 12}$ & ${\small 0}$ & ${\small 6}$ &  & $%
{\small 3}$ & ${\small 0}$ & ${\small 13}$ &  & ${\small 8}$ & ${\small 4}$
& ${\small 4}$ &  & ${\small 0}$ & ${\small 5}$ & ${\small 14}$ \\ \hline
{\small 2015} & ${\small 0.546}$ & ${\small 3}$ & ${\small 0}$ & ${\small 1}$
& ${\small 2.384}$ & ${\small 0}$ & ${\small 0}$ & ${\small 4}$ & ${\small %
0.803}$ & ${\small 4}$ & ${\small 0}$ & ${\small 0}$ & ${\small 5.869}$ & $%
{\small 0}$ & ${\small 1}$ & ${\small 6}$ \\ 
{\small 2016} & ${\small 0.360}$ & ${\small 4}$ & ${\small 0}$ & ${\small 2}$
& ${\small 1.358}$ & ${\small 0}$ & ${\small 0}$ & ${\small 4}$ & ${\small %
1.920}$ & ${\small 1}$ & ${\small 3}$ & ${\small 0}$ & ${\small 6.403}$ & $%
{\small 0}$ & ${\small 0}$ & ${\small 4}$ \\ 
{\small 2017} & ${\small 0.233}$ & ${\small 1}$ & ${\small 0}$ & ${\small 3}$
& ${\small 3.043}$ & ${\small 0}$ & ${\small 0}$ & ${\small 4}$ & ${\small %
0.429}$ & ${\small 0}$ & ${\small 0}$ & ${\small 4}$ & ${\small 6.320}$ & $%
{\small 0}$ & ${\small 0}$ & ${\small 4}$ \\ 
{\small 2018} & ${\small 0.459}$ & ${\small 4}$ & ${\small 0}$ & ${\small 0}$
& ${\small 3.463}$ & ${\small 3}$ & ${\small 0}$ & ${\small 1}$ & ${\small %
1.708}$ & ${\small 3}$ & ${\small 1}$ & ${\small 0}$ & ${\small 5.825}$ & $%
{\small 0}$ & ${\small 4}$ & ${\small 0}$%
\end{tabular}%
\bigskip

\newpage
\end{center}

\bigskip

\begin{center}
Table 8: Classification of the combination procedures according to their
forecast ability

\begin{tabular}{lcccccc}
& \multicolumn{2}{c}{$AVERAGE$} & \multicolumn{2}{c}{$S\_STC$} & 
\multicolumn{2}{c}{$CAS$} \\ \cline{2-7}
& $\#$ & $\%$ & $\#$ & $\%$ & $\#$ & $\%$ \\ \hline
Case 1 & $40$ & $32.8$ & $8$ & $17.8$ & $63$ & $52.1$ \\ 
Case 2 & $7$ & $5.7$ & $6$ & $13.3$ & $16$ & $13.2$ \\ 
Case 3 & $60$ & $49.2$ & $27$ & $60.0$ & $24$ & $19.8$ \\ 
Case 4 & $15$ & $12.3$ & $4$ & $8.9$ & $18$ & $14.9$ \\ \hline
Total & $122$ &  & $45$ &  & $121$ &  \\ \hline
\end{tabular}

{\small \#: Proportions (bias, variance and covariance) of the best MSFE
procedure included in specific cases.}\bigskip
\end{center}

\newpage


\newpage

\begin{thebibliography}{99}
\bibitem{} Aitchison, J. (1982) The statistical analysis of compositional
data (with discussion). \emph{J. R. Statist. Soc. B}, 44, 139--177.

\bibitem{} Aitchison, J. (1986) \emph{The statistical analysis of
compositional data}, Monographs on Statistics and Applied Probability.
London: Chapman \& Hall, (Reprinted in 2003 with additional material by The
Blackburn Press)

\bibitem{} Arroyo, A.S.M. and A. de Juan Fern{\'{a}}ndez (2014),
Split-then-Combine method for out-of-sample combinations of forecasts, \emph{%
J. Bus Adm Res}, 3 (1), pp. 19 - 37.

\bibitem{} Barrow, D.K. and Kourentzes, N. (2016). Distributions of
forecasting errors of forecast combinations: implications for inventory
management. \emph{International Journal of Production Economics, }177, pp.
24-33.

\bibitem{} Billheimer, D., P. Guttorp, and W. Fagan, (2001), Statistical
interpretation of species composition, \emph{Journal of the American
Statitictical Asociation}, 96 (456), 2001, pp. 1205--1214.

\bibitem{} van den Boogaart, K.G., R. Tolosana, and M. Bren (2009), \emph{%
Compositions: Compositional data analysis. R package}.

\bibitem{} Bujosa-Brun, M. A. Garc\'{\i}a-Ferrer, A. de Juan and A. Mart%
\'{\i}n Arroyo (2019): Evaluating early warning and coincident indicators of
business cycles using smooth trends, forthcoming \textit{Journal of
Forecasting. DOI: 10.1002/for.2601.}

\bibitem{} Diebold, F.X. and Mariano, R.S. (1995). Comparing predictive
accuracy. \emph{Journal of Business and Economic Statistics}, vol 13(3). pp.
253-263.

\bibitem{} Egozcue, J.J. and V. Pawlowsky-Glahn, (2005) Groups of parts and
their balances in compositional data analysis. \emph{Mathematical Geology},
37 (7), 2005, pp. 795--828.

\bibitem{} Federal Reserve Bank of Philadelphia (2018),. Survey of
Professional Forecasters documentation. August 29, 2018.

\bibitem{} Gabriel, K.R. (1971). The biplot---graphic display of matrices
with application to principal component analysis. \emph{Biometrika}, 58 (3),
pp. 453--467.

\bibitem{} Genre, V., Kenny, G., Meyler, A. and Timmermann, A. (2013).
Combining expert forecasts: Can anything beat the simple average?. \emph{%
International Journal of Forecasting,} 29, pp. 108-121.

\bibitem{} Giacomini, R. and White, H. (2006). Test of conditional
predictive ability, \emph{Econometrica}, 74(6), pp. 1545-1578.

\bibitem{} Lahiri, K., H. Peng and Y. Zhao (2017): Online learnig and
forecast combination in unbalanced panels. \emph{Econometric Reviews,}%
\textit{\ }vol. 36, pp. 257 - 288\textit{.}

\bibitem{} Jose, V. R. and R.L. Winkler (2008): Simple robust averages of
forecasts: Some empirical results. \emph{International Journal of Forecasting%
}\textit{, }24, 163 - 1 69

\bibitem{} Pawlowsky-Glahn, V. and Buccianti, A. (2011) \emph{Compositional
Data Analysis: Theory and Applications}. Chichester: Wiley.

\bibitem{} Poncela, P., J. Rodr{\'{\i}}guez, R. S{\'{a}}nchez-Mangas, and E.
Senra (2011), Forecast combination through dimension reduction techniques. 
\emph{International Journal of Forecasting}\textit{, }27, pp. 224 - 237.

\bibitem{} Smith, J. and Wallis, K.F. (2009), A simple explanation of the
forecast combination puzzle. \emph{Oxford bulletin of economics and
statistics}, 71(3), pp.331-355.

\bibitem{} Stock, J.H. and Watson, M. W. (2004). Combination forecasts of
output growth in a seven country dataset, \emph{Journal of Forecasting, }%
vol. 23, pp. 405-430.\bigskip
\end{thebibliography}
\end{document}